\documentclass[12pt,preprint]{emulateapj}

\begin{document}
\title{Very blue $UV$-continuum slopes $\beta$ of low luminosity
  $z\sim7$ galaxies from WFC3/IR: Evidence for extremely low
  metallicities?\altaffilmark{1}}
\author{R. J. Bouwens\altaffilmark{2,3},
  G. D. Illingworth\altaffilmark{2}, P. A. Oesch\altaffilmark{4},
  M. Trenti\altaffilmark{5}, M. Stiavelli\altaffilmark{6},
  C. M. Carollo\altaffilmark{4}, M. Franx\altaffilmark{3}, P. G. van
  Dokkum\altaffilmark{7}, I. Labb\'{e}\altaffilmark{8},
  D. Magee\altaffilmark{2}} \altaffiltext{1}{Based on observations
  made with the NASA/ESA Hubble Space Telescope, which is operated by
  the Association of Universities for Research in Astronomy, Inc.,
  under NASA contract NAS 5-26555. These observations are associated
  with programs \#11563, 9797.}  \altaffiltext{2}{UCO/Lick
  Observatory, University of California, Santa Cruz, CA 95064}
\altaffiltext{3}{Leiden Observatory, Leiden University, NL-2300 RA
  Leiden, Netherlands} \altaffiltext{4}{Institute for Astronomy, ETH
  Zurich, 8092 Zurich, Switzerland}
\altaffiltext{5}{University of Colorado, Center for Astrophysics and
  Space Astronomy, 389-UCB, Boulder, CO 80309, USA}
\altaffiltext{6}{Space Telescope Science Institute, Baltimore, MD
  21218, United States} \altaffiltext{7}{Department of Astronomy, Yale
  University, New Haven, CT 06520} \altaffiltext{8}{Carnegie
  Observatories, Pasadena, CA 91101, Hubble Fellow}

\begin{abstract}
We use the ultra-deep WFC3/IR data over the HUDF and the Early Release
Science WFC3/IR data over the CDF-South GOODS field to quantify the
broadband spectral properties of candidate star-forming galaxies at
$z\sim7$.  We determine the $UV$-continuum slope $\beta$ in these
galaxies, and compare the slopes with galaxies at later times to
measure the evolution in $\beta$.  For luminous $L_{z=3}^{*}$
galaxies, we measure a mean $UV$-continuum slope $\beta$ of
$-2.0\pm0.2$, which is comparable to the $\beta\sim-2$ derived at
similar luminosities at $z\sim5-6$.  However, for the lower luminosity
$0.1L_{z=3}^{*}$ galaxies, we measure a mean $\beta$ of $-3.0\pm0.2$.
This is substantially bluer than is found for similar luminosity
galaxies at $z$$\sim$4, just 800 Myr later, and even at $z$$\sim$5-6.
In principle, the observed $\beta$ of $-3.0$ can be matched by a very
young, dust-free stellar population, but when nebular emission is
included the expected $\beta$ becomes $\geq-2.7$.  To produce these
very blue $\beta$'s (i.e., $\beta\sim-$3), extremely low metallicities
and mechanisms to reduce the red nebular emission seem to be required.
For example, a large escape fraction (i.e., $f_{esc}\gtrsim0.3$) could
minimize the contribution from this red nebular emission.  If this is
correct and the escape fraction in faint $z\sim7$ galaxies is
$\gtrsim$0.3, it may help to explain how galaxies reionize the
universe.
\end{abstract}

\section{Introduction}

The spectral properties of high-redshift galaxies must undergo
dramatic changes at some point in the past, as the metallicities in
these systems drop to lower values and these systems become
progressively younger.  In the limit of low metallicities, gas is no
longer able to cool efficiently, likely resulting in massive extremely
low-metallicity (or Population III) stars whose hot atmospheres are
expected to result in a very hard $UV$-continuum spectrum and strong
HeII 1640 emission.  However, the strong UV flux coming from hot stars
is expected to be largely offset by the redder nebular continuum light
produced by the ionized gas surrounding these massive stars (e.g.,
Schaerer 2002).

The newly installed WFC3/IR camera on the Hubble Space Telescope
permits us to observe faint $z\gtrsim7$ galaxies $\gtrsim40\times$
more efficiently than before, providing us with our most detailed look
yet at the $UV$ light and spectral properties of $z\gtrsim7$ galaxies.
Already $\gtrsim$25 likely $z$$\sim$7-8 galaxies have been identified
in the early WFC3/IR data over the Hubble Ultra Deep Field (HUDF:
Oesch et al.\ 2009b; Bouwens et al.\ 2009b; McLure et al.\ 2009;
Bunker et al.\ 2009), and $\gtrsim$20 $z\sim7$ candidate galaxies in
the WFC3/IR Early Release Science (ERS) observations over the
CDF-South (R.J. Bouwens et al.\ 2009, in prep).

Here we take advantage of these early WFC3/IR observations to study
the spectral properties of candidate $z\sim7$ galaxies.  Our principal
focus will be on the slope of $z\sim7$ galaxy spectra in the
$UV$-continuum -- since this slope is the primary observable we can
derive from the available broadband imaging data with WFC3.  The
$UV$-continuum slope $\beta$ ($f_{\lambda}\propto \lambda^{\beta}$:
e.g., Meurer et al.\ 1999) provides us with a powerful constraint on
the age, metallicity, and dust content of high-redshift galaxies; it
has already been the subject of much study at $z\sim3-6$ (Lehnert \&
Bremer 2003; Stanway et al.\ 2005; Yan et al.\ 2005; Bouwens et
al.\ 2006; Hathi et al.\ 2008; see Bouwens et al. 2009a for a
systematic study at $z\sim2-6$) and even at $z\sim7$ (Gonzalez et
al.\ 2009) using NICMOS data.

Throughout this work, we quote results in terms of the luminosity
$L_{z=3}^{*}$ Steidel et al.\ (1999) derived at $z\sim3$:
$M_{1700,AB}=-21.07$.  We refer to the HST F606W, F775W, F850LP,
F105W, F125W, and F160W bands as $V_{606}$, $i_{775}$, $z_{850}$,
$Y_{105}$, $J_{125}$, and $H_{160}$, respectively, for simplicity.
Where necessary, we assume $\Omega_0 = 0.3$, $\Omega_{\Lambda} = 0.7$,
$H_0 = 70\,\textrm{km/s/Mpc}$.  All magnitudes are in the AB system
(Oke \& Gunn 1983).

\section{Observational Data}

Our primary dataset is the ultra deep WFC3/IR data over the HUDF and
the wide-area WFC3/IR ERS observations over the CDF-South (PI
O'Connell: GO 11359).  The HUDF data permit us to identify lower
luminosity $z\sim7$ galaxies and quantify their properties (e.g.,
Oesch et al.\ 2009b), while the ERS data permit us to do the same for
more luminous $z\sim7$ galaxies.

For our HUDF dropout selections, we make use of the v1.0 reductions of
the HUDF ACS data (Beckwith et al.\ 2006) rebinned on a 0.06$''$-pixel
scale, and our own reduction of the HUDF09 WFC3/IR data over the HUDF
(Bouwens et al.\ 2009b; Oesch et al.\ 2009b).  The optical ACS imaging
over the HUDF reach to 29.4, 29.8, 29.7, and 29.0 AB mag (5$\sigma$:
0.35$''$-diameter apertures) in the $B_{435}$, $V_{606}$, $i_{775}$,
$z_{850}$ bands, respectively, while the near-IR WFC3/IR data reach to
28.8, 28.8, and 28.8 ($5\sigma$: 0.35$''$ apertures) in the $Y_{105}$,
$J_{125}$, and $H_{160}$ bands, respectively.  The PSF FWHMs are
$\sim$0.10$''$ for the ACS data and $\sim$0.16$''$ for the WFC3/IR
data.

For our dropout selections over the WFC3/IR ERS fields, we make use of
our own reductions of the available ACS/WFC data over the GOODS fields
(Bouwens et al.\ 2006, 2007).  Our reductions reach to 28.0, 28.2,
27.5, and 27.4 in the $B_{435}$, $V_{606}$, $i_{775}$, $z_{850}$,
respectively ($5\sigma$: $0.35''$ apertures).  This is similar to the
GOODS v2.0 reductions, but reach $\sim$0.1-0.3 mag deeper in the
$z_{850}$-band due to the inclusion of the SNe follow-up data (Riess
et al.\ 2007).  Our reductions of the WFC3/IR ERS data was performed
using the same procedures as we used on the HUDF09 WFC3/IR data.
These data reach to 27.7 and 27.4 in the $J_{125}$ and $H_{160}$
bands, respectively ($5\sigma$: $0.35''$ apertures).

\section{Results}

\subsection{Catalog Creation} 

Our procedure for source detection and photometry is similar to that
used in previous studies by our team (e.g., Bouwens et al.\ 2007,
2009a) and relies upon the SExtractor software (Bertin \& Arnouts
1996) run in double image mode.  Source detection is performed off the
square root of $\chi^2$ image (Szalay et al.\ 1999: similar to a
coadded image) constructed from all images redward of the Lyman break.
Colors are measured in small scalable apertures with a Kron (1980)
parameter of 1.2 (typically $\sim$0.4$''$-diameter apertures).  Fluxes
measured in these small scalable apertures are then corrected
(typically by $\sim$0.4 mag) to total magnitudes using the additional
flux in a larger scalable aperture (Kron parameter of 2.5).

\begin{figure}
\epsscale{1.18}
\plotone{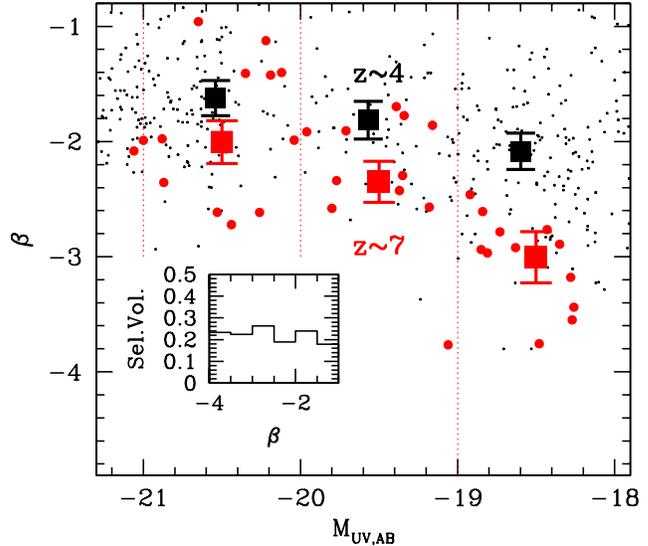}
\caption{$UV$-continuum slope $\beta$ versus absolute $UV$ magnitude.
  The red circles show the $\beta$ determinations and absolute
  magnitudes (derived from the $J_{125}-H_{160}$ colors and
  $\frac{1}{2}(J_{125}+H_{160})_{AB}$ magnitudes, respectively) for
  individual $z\sim7$ galaxies in our HUDF09 and ERS selections.  The
  large red squares (with $1\sigma$ error bars) represent the mean
  values in 1-mag bins (\textit{dotted lines}).  The black points
  correspond to the $\beta$ determinations at $z\sim4$ (Bouwens et
  al.\ 2009a).  The $\beta$'s for lower luminosity galaxies at
  $z\sim7$ (red circles) are much bluer (by $\Delta \beta$ of $\sim$1)
  than those derived at $z\sim4$ (a $4\sigma$ difference).
  \textit{(inset)} The relative volumes available for selecting
  galaxies with various $\beta$'s using our HUDF09 $z$-dropout
  criterion (see \S3.4)).  These volumes do not depend significantly on
  $\beta$, demonstrating that the blue $\beta$'s observed for faint
  $z\sim7$ galaxies is not a selection effect.\label{fig:colmag}}
\end{figure}

\subsection{$z\sim7$ $z$-dropout selection}

For our lower-luminosity $z\sim7$ $z$-dropout selection, we use a
criterion very similar that used by Oesch et al.\ (2009b) over the
HUDF09 WFC3/IR field:
\begin{displaymath}
(z_{850}-Y_{105}>0.8) \wedge (Y_{105}-J_{125} < 0.8)
\end{displaymath}
Over the ERS fields, no deep WFC3 $Y_{105}$-band coverage is
available, so we make use of the following criterion (R.J. Bouwens
2009, in prep):
\begin{displaymath}
(z_{850}-J_{125}>0.8) \wedge (J_{125}-H_{160} < 0.3)
\end{displaymath}
Sources are required to be undetected ($<2\sigma$) in all bands
blueward of the break (and also undetected [$<1.5\sigma$] in no $>1$
bands) to ensure that our selections are largely free of lower
redshift contaminants.  Sources must be detected at $\geq$5.5$\sigma$
in the $J_{125}$ band to ensure that $z\sim7$ candidates in our sample
correspond to real sources.

\begin{figure}
\epsscale{1.16} \plotone{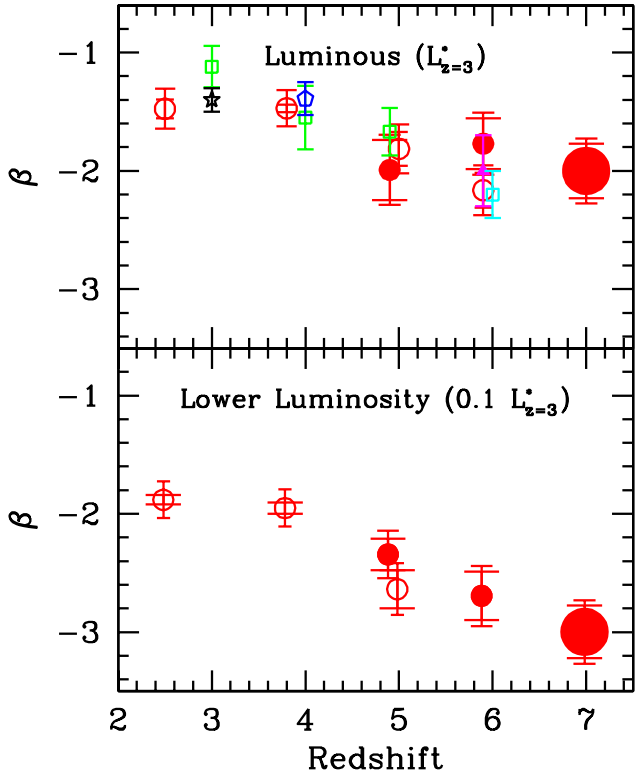}
\caption{Mean $UV$-continuum slope measured for galaxies at $z\sim7$
  (large red circles) versus galaxies of similar luminosities at lower
  redshifts.  The top panel shows the evolution in this slope for
  galaxies with a $UV$ luminosity of $L_{z=3}^{*}$ ($-$21 AB mag) and
  the bottom panel shows this evolution for $UV$ luminosities of
  $0.1L_{z=3}^{*}$ ($-$18.5 AB mag).  The determinations of Bouwens et
  al.\ (2009a) based upon ACS+NICMOS data are shown with the open red
  circles.  The short error bars are the random errors while the
  longer error bars also include possible systematic errors (e.g., see
  Bouwens et al.\ 2009a).  We also show $\beta$ determinations by
  Stanway et al.\ (2005: \textit{cyan square}), Ouchi et al.\ (2004:
  \textit{blue pentagon}), Adelberger \& Steidel (2000: \textit{black
    star}), and Hathi et al.\ (2008: \textit{green open squares}).
  Note the dramatic change in $\beta$ in the $\sim$800 Myr from
  $z\sim7$ to $z$$\sim$4 for lower luminosity
  galaxies.\label{fig:mcolor}}
\end{figure}

\subsection{$UV$-continuum slope Measurements}

The $UV$-continuum slopes $\beta$ we estimate for sources in our
$z\sim7$ $z$-dropout sample are derived from the broadband color
$J_{125}-H_{160}$ as
\begin{equation}
\beta = 4.29 (J_{125}-H_{160}) - 2.00\label{eq:form}
\end{equation}
The $J_{125}$ and $H_{160}$ bands here probe rest-frame $\sim$1550\AA
and $\sim$1940\AA, respectively, for the typical $z\sim7$ $z$-dropout
candidate in our sample, and are not affected by Ly$\alpha$ emission
or IGM absorption (in contrast to the $Y_{105}$-band).  The above
equation is derived assuming a flat spectrum source with no absorption
lines (found to work very well by Meurer et al.\ 1999).

In Figure~\ref{fig:colmag}, we show the $\beta$ determinations for
$z\sim7$ $z$-dropout candidates versus luminosity.  Both our
ultra-deep HUDF09 and wide-area ERS selections are included, as are
the $z\sim4$ selections from Bouwens et al.\ (2009a).  The trend of
$\beta$ with luminosity is illustrated with the large squares.  The
difference between $z\sim7$ and $z\sim4$ is striking.  

Use of the photometry from Oesch et al.\ (2009b), McLure et
al.\ (2009), or Bunker et al.\ (2009) for our $z\sim7$ sample yield
very consistent colors and similarly blue (or even bluer) $\beta$'s.
In addition, use of a consistent $0.7''$-diameter aperture for the
color measurements yield equally blue $\beta$'s (albeit with larger
random uncertainties, due to the use of larger apertures).

\subsection{Consideration of Possible Selection Biases}

The distribution of $\beta$'s derived for sources in our $z\sim7$
sample can be affected by the selection process.  This effect is seen
in lower redshift samples, where galaxies with bluer $\beta$'s have
larger selection volumes than galaxies with redder $\beta$'s.  To
determine the importance of such effects, we constructed models with
various $\beta$ distributions, added galaxies with these distributions
to the observational data, and then attempted to reselect these
galaxies using the selection criteria in \S3.2 to estimate the
effective selection volume versus $\beta$.  We modelled the
pixel-by-pixel profiles of the sources using similar luminosity
$z\sim4$ $B$-dropouts from the Bouwens et al.\ (2007) HUDF sample, but
scaled in size (physical) as $(1+z)^{-1}$ to match the observed
size-redshift relationship (Oesch et al.\ 2009c; Ferguson et
al.\ 2004; Bouwens et al.\ 2004).

The selection volumes derived from these simulations are shown in the
inset to Figure~\ref{fig:colmag}.  Encouragingly, these volumes show
only a modest dependence upon the input $\beta$ for the range
$-4<\beta<-1$, and so selection biases are minimal.

\subsection{Low-redshift Comparison Samples}

To interpet the $\beta$'s we derive from our $z\sim7$ $z$-dropout
selections, we compare them with the $\beta$'s found for similar
luminosity galaxies at $z\sim4-6$.  Bouwens et al.\ (2009a) provide
determinations of these slopes as a function of luminosity at
$z\sim2-6$.  The $\beta$ measurements at $z\sim4$ are also shown in
Figure~\ref{fig:colmag} for comparison.

We can take advantage of the very deep, high-quality WFC3/IR data to
obtain self-consistent determinations of these slopes at $z$$\sim$5-6.
The $z\sim5$ $V$ and $z\sim6$ $i$-dropouts over the HUDF09 and GOODS
ERS data are selected in the same way as in Bouwens et al.\ (2007: see
also Giavalisco et al.\ 2004; Beckwith et al.\ 2006), except that at
$z\sim6$ we also require galaxies to satisfy the criterion
$(z_{850}-J_{125}<0.6)$.  At $z\sim5$ and $z\sim6$, the $\beta$'s are
estimated based upon the $z_{850}-(Y_{105}+J_{125})/2$ and
$Y_{105}-(2J_{125}+H_{160})/3$ colors, respectively, which are a good
match in rest-frame wavelength to the $J_{125}-H_{160}$ colors used to
estimate $\beta$ at $z\sim7$.  The conversion formulas we use are
\begin{eqnarray}
\beta = 4.07 (z_{850}-(Y_{105}+J_{125})/2) - 2.00~~&\textrm{($z\sim5$)} \\
\beta = 3.78 (Y_{105}-(2J_{125}+H_{160})/3) - 2.00~~&\textrm{($z\sim6$)}
\label{eq:beta}
\end{eqnarray}
In Figure~\ref{fig:mcolor}, we plot these $\beta$ determinations as a
function of redshift, for both luminous $L^*$ galaxies and lower
luminosity galaxies.

\subsection{$z\sim8$ Comparison Sample}

Faint $z\sim8$ $Y_{105}$-dropout candidates in the HUDF09 WFC3/IR data
(Bouwens et al.\ 2009b) also have very blue $J_{125}-H_{160}$ colors,
again suggesting $\beta$'s of $\sim-3$.  We decided not to consider
this sample here because the $J_{125}$-band flux can be affected by
Lyman series absorption and Ly$\alpha$ emission at $z>8.1$, and hence
the results would be less robust.

\section{Discussion}

Before discussing the extraordinarily blue $UV$-continuum slopes
$\beta\sim-3$ found for lower luminosity galaxies at $z\sim7$, we
first consider the relatively luminous $L_{z=3}^{*}$ galaxy
candidates.  These sources have measured $\beta$'s of $-2$, which is
similar to that observed for luminous galaxies at $z\sim3-6$
(Figure~\ref{fig:mcolor}).  Such slopes can be fit by a moderately
young, subsolar ($0.2\,Z_{\odot}$) stellar population, with a maximum
$E(B-V)$ of $\sim$0.05 (Calzetti et al.\ 2000 extinction law: see also
Bouwens et al.\ 2009a).

The lower luminosity ($0.1L_{z=3}^{*}$) galaxy candidates at $z\sim7$,
by contrast, have observed $\beta$'s of $-3.0\pm0.2$.  This is much
bluer than for luminous $z\sim7$ galaxies (by $\Delta \beta\sim1$) and
also much bluer than is found at $z\sim5-6$
(Figures~\ref{fig:colmag}-\ref{fig:mcolor}).  This makes these
galaxies of great interest since they are likely to be even younger,
more metal poor, and dust-free than any galaxies known.

This is not a selection effect (\S3.4), and cannot be attributed to
Ly$\alpha$ emission contributing to the broadband fluxes.  Ly$\alpha$
does not move into the $J_{125}$-band (used to estimate $\beta$) until
$z\gtrsim8.1$, and $\lesssim10$\% of our $z$-dropout sample extends to
$z>8$.

An AGN contribution seem similarly unlikely, given the rarety of AGN
signatures in faint Lyman-Break and Ly$\alpha$-emitter samples (e.g.,
Nandra et al.\ 2002; Ouchi et al.\ 2008).

What then is the explanation for the very blue $\beta$'s?  We explore
several possibilities:

\begin{figure}
\epsscale{1.16}
\plotone{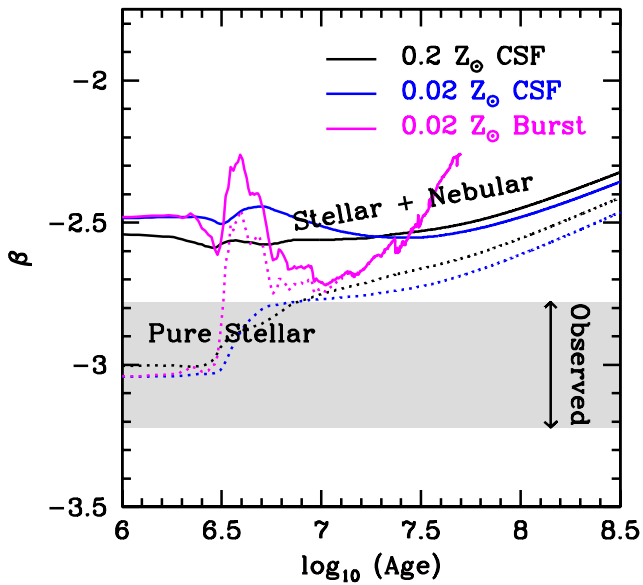}
\caption{$UV$-continuum slope $\beta$ we would expect for $z\sim7$
  galaxies as a function of age for constant star formation (CSF)
  models and an instantaneous burst (Schaerer 2002).  The gray band
  denotes the observed mean $\beta$ and its uncertainty.  The slopes
  $\beta$ derived from the stellar light (dotted lines) and the
  stellar + nebular light (solid lines) are shown.  While it is in
  principle possible to obtain $\beta$'s of $-3$ with standard low
  metallicity ($\geq$0.02 $Z_{\odot}$) models, including the nebular
  emission associated with hot stars make the predicted $\beta$'s
  $\geq-2.7$ and hence too red.\label{fig:dbeta}}
\end{figure}

\begin{figure}
\epsscale{1.16}
\plotone{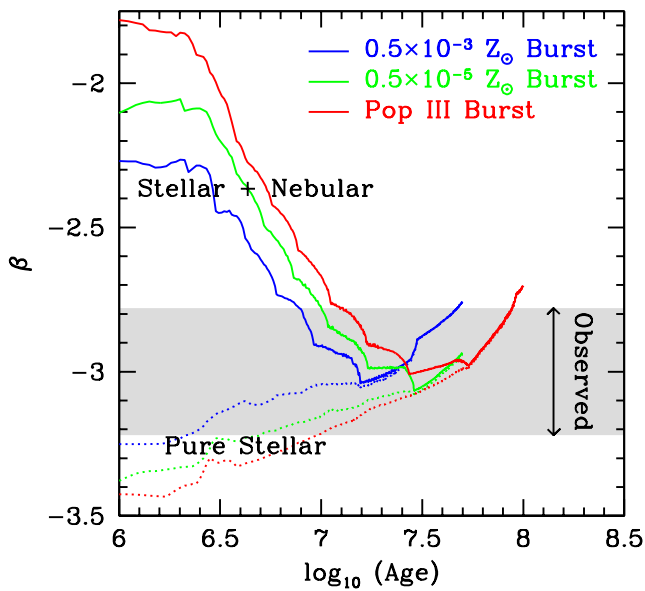}
\caption{$UV$-continuum slope $\beta$ Schaerer (2003) calculated as a
  function of age for instantaneous burst models for different
  metallicities.  Both the slopes $\beta$ derived from the stellar
  light (dotted lines) and the stellar + nebular light (solid lines)
  are shown.  The pure stellar light (dotted lines) has very blue
  $UV$-continuum slopes $\beta$'s ($\beta\lesssim-3$) for all the low
  metallicity cases considered here.  Changing the IMF does not appear
  to change the conclusion here in any significant way.  Of course,
  these same hot low metallicity stars also ionize the gas around
  them, thus producing a substantial amount of redder nebular
  continuum light.  This makes the total SED of a galaxy much redder
  in general, and in the calculations by Schaerer (2003) shown here,
  $\beta$ never becomes bluer than $-$3.0.
\label{fig:dbeta2}}
\end{figure}

\subsection{Standard stellar population models} 

A first question is whether it is possible to obtain a $\beta$ of
$\sim-3$ using standard stellar population models (e.g., Leitherer et
al.\ 1999; Bruzual \& Charlot 2003).  The answer is that it is
possible, but only for very young ($<$5 Myr) star-forming systems (see
Figure~\ref{fig:dbeta}).  However, to do so ignores the nebular
continuum emission from the ionized gas around the young stars.
Including this nebular continuum emission can redden the observed
$\beta$ by as much as $\Delta\beta\sim$0.5.

Figure~\ref{fig:dbeta} also shows the $\beta$'s predicted for several
low metallicity (0.02 $Z_{\odot}$ and 0.2 $Z_{\odot}$) starbursts, as
a function of age for the Schaerer (2002) stellar population models
(which -- like \textit{Starburst99} Leitherer et al.\ 1999 -- include
a nebular contribution).  In the best cases, the models predict
$\beta$'s as blue as $-2.7$, which is redder than what we observe in
our faint samples.\footnote{To match the wavelength baseline for
  $\beta$ measured here, we added $-$0.1 to the $\beta$'s
  (1300-1800\AA baseline) tabulated in the Schaerer (2002, 2003).}
This would suggest that lower metallicities are needed since the
standard Leitherer et al.\ (1999) or Bruzual \& Charlot (2003) stellar
population models do not produce blue enough colors to match those
found in our lower luminosity $z\sim7$ sample.  Of course, the
significance of this result is only modest ($<1.3\sigma$), but the
similarly blue colors observed for z$\sim$8 selections (Section 3.6)
suggest that we may want to take this finding at face value.

\subsection{Extremely low metallicities ($\leq10^{-3}\,\,Z_{\odot}$)}

In Figure~\ref{fig:dbeta2} we present the $\beta$'s predicted by the
Schaerer (2003) stellar population models (which conveniently provide
predictions at extremely low metallicities) as a function of age for
instantaneous bursts assuming a metallicity of $0.5\times10^{-3}
Z_{\odot}$, $0.5\times10^{-5} Z_{\odot}$, and zero (population
III).\footnote{The use of instantaneous burst models allows us to
  explore the most extreme cases.  Other models ($\tau$, inverse
  $\tau$, CSF) produce comparable $\beta$'s.}  From the figure, it is
clear that the nebular component contributes significantly to the
total light output from $\lesssim10^{-3}$ $Z_{\odot}$ stellar
populations.

While initially somewhat red due to the nebular contribution, the
predicted $\beta$'s for these ultra-low metallicity models become much
bluer at ages $>10$ Myr, eventually reaching $\beta$'s of $-$3.  Such
metallicities and ages are not necessarily unreasonable for lower
luminosity galaxies at $z\sim7$, and therefore at least one possible
explanation for the very blue $\beta$'s in our selections is that the
metallicities for galaxies in our sample may be
$\lesssim10^{-3}\,Z_{\odot}$.

The above explanation may explain some of the very blue galaxies in
our selection, but given that $\beta\sim-3$ only for a limited period
(10-30 Myr after a burst), it seems unlikely to be the general
explanation (unless updated models revise the theoretical SEDs).

\subsection{Top-heavy IMF}

One seemingly attractive explanation for the very blue $\beta$'s
observed is through a top-heavy IMF, since galaxy stellar populations
would be weighted towards massive, blue stars.  The difficulty with
this explanation is that these same massive stars are extraordinarily
efficient at ionizing the gas around them -- resulting in substantial
nebular emission and leaving the galaxy with a net $\beta$ no bluer
(and likely redder) than the young ($<$1 Myr) bursts shown in
Figures~\ref{fig:dbeta}-\ref{fig:dbeta2} (see also discussion in
Leitherer \& Heckman 1995).

\subsection{Minimizing nebular emission} 

A significant obstacle to matching the very blue $\beta$'s observed is
the red nebular emission associated with hot, ionizing stars.  The
nebular contribution could be reduced in a number of ways, by changes
to the ionization parameter, metallicity, geometry, etc.  Assessing
the impact of such changes would benefit from further detailed
modelling.  However, we should emphasize that regardless of changes to
the nebular contribution very low metallicity models (or very young
ages) appear to be needed to match the very blue $\beta$'s observed.

\begin{figure}
\epsscale{1.16}
\plotone{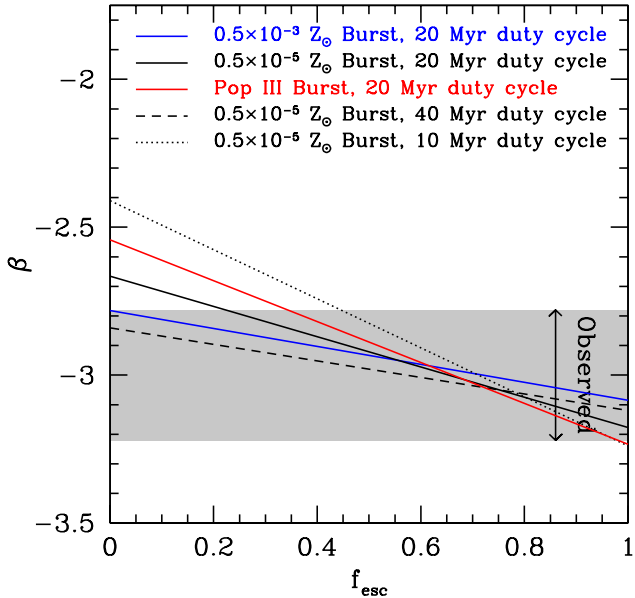}
\caption{Mean $UV$-continuum slopes $\beta$'s predicted for
  high-redshift galaxy samples versus the escape fraction $f_{esc}$.
  The predictions are made using the Schaerer (2003) stellar
  population models (including nebular emission) for metallicities of
  $<10^{-3}$ $Z_{\odot}$.  Galaxies are assumed to be observed at some
  random point during their star formation histories and to experience
  instantaneous bursts of star formation every 10-40 Myr (the dashed,
  solid, and dotted lines give the average $\beta$ for the first 10,
  20, and 40 Myr, respectively, of the instantaneous burst).  For
  $f_{esc}=1$, all of the ionizing radiation from a galaxy escapes
  into the IGM and hence does not contribute to ionizing the gas
  within a galaxy (and hence the contribution from nebular continuum
  emission to the total light is minimal).  On the other hand, for
  $f_{esc}\sim0$ (preferred in some simulations: e.g., Gnedin et
  al.\ 2008), the ionizing radiation from the hot stars does not make
  it out of galaxies -- resulting in substantial nebular continuum
  emission (see Figure~\ref{fig:dbeta2}) and hence a much redder
  $\beta$.  Perhaps the very blue $\beta$'s observed (\textit{shaded
    gray region}) could indicate that the escape fraction is larger at
  $f_{esc}\gtrsim0.3$ than what has commonly been considered?
\label{fig:escapef}}
\end{figure}

\subsection{Changes in Escape Fraction?} 

One possibility that we have explored to reduce the nebular emission
contribution is if the ionizing radiation leaks directly into the IGM
(e.g., due to the effect of SNe on the galaxies' ISM, possibly from a
top-heavy IMF: Trenti \& Shull 2009).  We consider such a possibility
schematically in Figure~\ref{fig:escapef}, showing the time-averaged
$\beta$'s expected for stellar populations of various metallicities as
a function of the escape fraction $f_{esc}$ of ionizing photons into
the IGM.  For $f_{esc}$ of unity, we simply recover the $\beta$'s from
the pure stellar SEDs and for $f_{esc}$ of zero, we recover the
stellar + nebular SEDs.  For fractional $f_{esc}$, we interpolate
between the two extremes.  The time-averaged $\beta$'s used for
Figure~\ref{fig:escapef} is based upon those presented in
Figure~\ref{fig:dbeta2}.

Comparing the predicted $\beta$'s with that observed (grey-shaded
region: Figure~\ref{fig:escapef}), we see that $f_{esc}\gtrsim$0.3
would permit us to easily match the observations.  Such an escape
fraction is signiﬁcantly higher than the $\sim$10\% frequently assumed
in calculations assessing the sufficiency of galaxies to reionize the
universe.  Since most of the luminosity density at $z>7$ comes from
low luminosity galaxies, the estimated number of ionizing photons in
the $z>7$ universe could increase by factors of $\gtrsim$3. Such a
large change could provide the needed photons to reionize the
universe, providing a resolution to the current debate (e.g., Bouwens
et al. 2008; Oesch et al. 2009a, 2009b; McLure et al. 2009; Bunker et
al. 2009; Gonzalez et al. 2009; Ouchi et al.\ 2009; Pawlik et
al.\ 2009).

\section{Summary}

The strikingly blue UV-continum slopes $\beta$'s seen at lower
luminosities in candidate $z\sim7$ galaxies (also apparent in the
Bouwens et al.\ 2009b $z\sim8$ sample) indicate that we are now
beginning to explore a regime where the nature of galaxies and their
stellar populations are undergoing a dramatic change.  These results
raise many as yet unanswered questions, but could be heralding the
transition from the first stars and youngest objects within the first
400 Myr at $z\gtrsim$10 to the galaxies that dominate the universe for
the next 2 Gyr and provide clues as to the source of photons that
reionize the universe.

\acknowledgements

We thank Daniel Schaerer for helpful conversations.  We are grateful
to all those at NASA, STScI and throughout the community who have
worked so diligently to make Hubble the remarkable observatory that it
is today.  We acknowledge the support of NASA grant NAG5-7697 and NASA
grant HST-GO-11563.01.


\begin{thebibliography}{} 
\bibitem[Adelberger \& Steidel(2000)]{2000ApJ...544..218A} Adelberger, 
K.~L.~\& Steidel, C.~C.\ 2000, \apj, 544, 218
\bibitem[Beckwith et al.(2006)]{2006AJ....132.1729B} Beckwith, S.~V.~W., et 
al.\ 2006, \aj, 132, 1729
\bibitem[Bertin and Arnouts (1996)]{1996A&AS..117..393B} Bertin, E.\ and 
Arnouts, S.\ 1996, \aaps, 117, 39
\bibitem[Bouwens et al.(2004)]{2004ApJ...611L...1B} Bouwens, R.~J., 
Illingworth, G.~D., Blakeslee, J.~P., Broadhurst, T.~J., 
\& Franx, M.\ 2004, \apjl, 611, L1
\bibitem[Bouwens et al. (2006)]{2006Bouwens} Bouwens, R.J., Illingworth,
G.D., Blakeslee, J.P., \& Franx, M.  2006, \apj, 653, 53 
\bibitem[Bouwens et al.(2007)]{2007ApJ...670..928B} Bouwens, R.~J., 
Illingworth, G.~D., Franx, M., \& Ford, H.\ 2007, \apj, 670, 928
\bibitem[Bouwens et al.(2008)]{2008ApJ...686..230B} Bouwens, R.~J., 
Illingworth, G.~D., Franx, M., \& Ford, H.\ 2008, \apj, 686, 230 
\bibitem[Bouwens et al.(2009)]{2009ApJ...705..936B} Bouwens, R.~J., et
  al.\ 2009a, \apj, 705, 936
\bibitem[Bouwens et al. (2009b)]{eee2} Bouwens, R.J., et al.\ 2009b,
  \apj, in press, arXiv:0909.1803
\bibitem[Bruzual 
\& Charlot(2003)]{2003MNRAS.344.1000B} Bruzual, G., \& Charlot, S.\ 2003, \mnras, 344, 1000 
\bibitem[Bunker et al. (2009)]{eee} Bunker, A., et al.\ 2009, \mnras,
  in press, arXiv:0909.2255
\bibitem[Calzetti et al.(2000)]{2000ApJ...533..682C} Calzetti, D., Armus, 
L., Bohlin, R.~C., Kinney, A.~L., Koornneef, J., 
\& Storchi-Bergmann, T.\ 2000, \apj, 533, 682 
\bibitem[Ferguson et al.(2004)]{2004ApJ...600L.107F} Ferguson, H.~C.~et 
al.\ 2004, \apjl, 600, L107
\bibitem[Giavalisco et al.(2004)]{2004ApJ...600L..93G} Giavalisco, M., et 
al.\ 2004a, \apjl, 600, L93
\bibitem[Gnedin et al.(2008)]{2008ApJ...672..765G} Gnedin, N.~Y., Kravtsov, 
A.~V., \& Chen, H.-W.\ 2008, \apj, 672, 765 
\bibitem[Gonzalez et al. (2009)]{eee} Gonzalez, V., Labbe, I., Bouwens,
  R., Illingworth, G., Franx, M., Kriek, M., Brammer, G.  2009, \apj,
  submitted, arXiv:0909.3517
\bibitem[Hathi et al.(2008)]{2008ApJ...673..686H} Hathi, N.~P., Malhotra, 
S., \& Rhoads, J.~E.\ 2008, \apj, 673, 686
\bibitem[Kron (1980)]{kron} Kron, R. G. 1980, \apjs, 43, 305
\bibitem[Lehnert \& Bremer(2003)]{2003ApJ...593..630L} Lehnert, M.~D.~\& 
Bremer, M.\ 2003, \apj, 593, 630
\bibitem[Leitherer 
\& Heckman(1995)]{1995ApJS...96....9L} Leitherer, C., \& Heckman, T.~M.\ 1995, \apjs, 96, 9 
\bibitem[Leitherer et al.(1999)]{1999ApJS..123....3L} Leitherer, C., et 
al.\ 1999, \apjs, 123, 3 
\bibitem[McLure et al. (2009)]{eee} McLure, R., et al.\ 2009, \mnras,
  in press, arXiv:0909.2437
\bibitem[Meurer et al.(1999)]{1999ApJ...521...64M} Meurer, G.~R., Heckman, 
T.~M., \& Calzetti, D.\ 1999, \apj, 521, 64 
\bibitem[Nandra et al.(2002)]{2002ApJ...576..625N} Nandra, K., Mushotzky, 
R.~F., Arnaud, K., Steidel, C.~C., Adelberger, K.~L., Gardner, J.~P., 
Teplitz, H.~I., \& Windhorst, R.~A.\ 2002, \apj, 576, 625
\bibitem[Oesch et al.(2009)]{2009ApJ...690.1350O} Oesch, P.~A., et al.\ 
2009a, \apj, 690, 1350 
\bibitem[Oesch et al. (2009)]{eee} Oesch, P.A., et al.\ 2009b, \apj,
  in press, arXiv:0909.1806
\bibitem[Oesch et al. (2009)]{eee} Oesch, P.A., et al.\ 2009c, \apj,
  in press, arXiv:0909.5183
\bibitem[Oke \& Gunn(1983)]{1983ApJ...266..713O} Oke, J.~B., \& Gunn, 
J.~E.\ 1983, \apj, 266, 713 
\bibitem[Ouchi et al.(2004)]{2004ApJ...611..660O} Ouchi, M., et al.\ 2004, 
\apj, 611, 660
\bibitem[Ouchi et al.(2008)]{2008ApJS..176..301O} Ouchi, M., et al.\ 2008, 
\apjs, 176, 301 
\bibitem[Ouchi et al.(2009)]{2009arXiv0908.3191O} Ouchi, M., et al.\ 2009, 
\apj, 706, 1136
\bibitem[Pawlik et al.(2009)]{2009MNRAS.394.1812P} Pawlik, A.~H., Schaye, 
J., \& van Scherpenzeel, E.\ 2009, \mnras, 394, 1812
\bibitem[Riess et al.(2007)]{2007ApJ...659...98R} Riess, A.~G., et al.\ 
2007, \apj, 659, 98 
\bibitem[Schaerer(2002)]{2002A&A...382...28S} Schaerer, D.\ 2002, \aap, 382, 28 
\bibitem[Schaerer(2003)]{2003A&A...397..527S} Schaerer, D.\ 2003, \aap, 397, 527 
\bibitem[Stanway et al.(2005)]{2005MNRAS.359.1184S} Stanway, E.~R., 
McMahon, R.~G., \& Bunker, A.~J.\ 2005, \mnras, 359, 1184
\bibitem[Steidel et al.\ (1999)]{1999ApJ...519....1S} Steidel, C.\ C.,
Adelberger, K.\ L., Giavalisco, M., Dickinson, M.\ and Pettini, M.\ 1999,
\apj, 519, 1
\bibitem[Szalay et al.(1999)]{1999AJ....117...68S} Szalay, A.~S.,
Connolly, A.~J., \& Szokoly, G.~P.\ 1999, \aj, 117, 68
\bibitem[Trenti 
\& Shull(2009)]{2009arXiv0905.4505T} Trenti, M., \& Shull, M.\ 2009, \apj, submitted, arXiv:0905.4505 
\bibitem[Yan et al.(2005)]{yan5} Yan, H., et al.\ 2005, \apj, 634, 109
\end{thebibliography}
\end{document}